\documentclass[pdflatex,sn-mathphys-num]{sn-jnl}

\usepackage{graphicx}%
\usepackage{multirow}%
\usepackage{amsmath,amssymb,amsfonts}%
\usepackage{amsthm}%
\usepackage{mathrsfs}%
\usepackage[title]{appendix}%
\usepackage{xcolor}%
\usepackage{textcomp}%
\usepackage{manyfoot}%
\usepackage{booktabs}%
\usepackage{algorithm}%
\usepackage{algorithmicx}%
\usepackage{algpseudocode}%
\usepackage{listings}%

\theoremstyle{thmstyleone}%
\theoremstyle{thmstyletwo}%

\theoremstyle{thmstylethree}%

\begin{document}

\title[Article Title]{Hardware acceleration for ultra-fast Neural Network training on FPGA for MRF map reconstruction}


\author[1,2]{\fnm{Mattia} \sur{Ricchi}}\email{mattia.ricchi@phd.unipi.it}
\equalcont{These authors contributed equally to this work.}

\author[2]{\fnm{Fabrizio} \sur{Alfonsi}}\email{fabrizio.alfonsi@cern.ch}
\equalcont{These authors contributed equally to this work.}

\author[3]{\fnm{Camilla} \sur{Marella}}\email{camilla.marella2@unibo.it}

\author[4]{\fnm{Marco} \sur{Barbieri}}\email{mb7@stanford.edu}

\author[5]{\fnm{Alessandra} \sur{Retico}}\email{alessandra.retico@pi.infn.it}

\author*[3]{\fnm{Leonardo} \sur{Brizi}}\email{leonardo.brizi2@unibo.it}

\author[2,3]{\fnm{Alessandro} \sur{Gabrielli}}\email{alessandro.gabrielli@unibo.it}
\equalcont{These authors contributed equally to this work.}

\author[2,3]{\fnm{Claudia} \sur{Testa}}\email{claudia.testa@unibo.it}
\equalcont{These authors contributed equally to this work.}

\affil*[1]{\orgdiv{Department of Computer Sciences}, \orgname{University of Pisa}, \orgaddress{\street{Largo Bruno Pontecorvo}, \city{Pisa}, \postcode{56127}, \country{Italy}}}

\affil[2]{\orgname{National Institute of Nuclear Physics (INFN), Division of Bologna}, \orgaddress{\street{Viale Berti Pichat}, \city{Bologna}, \postcode{40127}, \country{Italy}}}

\affil[3]{\orgdiv{Department of Physics and Astronomy}, \orgname{University of Bologna}, \orgaddress{\street{Viale Berti Pichat}, \city{Bologna}, \postcode{40127}, \country{Italy}}}

\affil[4]{\orgdiv{Department of Radiology}, \orgname{Stanford University}, \orgaddress{\city{Stanford}, \postcode{CA}, \country{USA}}}

\affil[5]{\orgname{National Institute of Nuclear Physics (INFN), Division of Pisa}, \orgaddress{\street{Largo Bruno Pontecorvo}, \city{Pisa}, \postcode{56127}, \country{Italy}}}


\abstract{Magnetic Resonance Fingerprinting (MRF) is a fast quantitative MR Imaging technique that provides multi-parametric maps with a single acquisition. Neural Networks (NNs) accelerate reconstruction but require significant resources for training. We propose an FPGA-based NN for real-time brain parameter reconstruction from MRF data. Training the NN takes an estimated 200 seconds, significantly faster than standard CPU-based training, which can be up to 250 times slower. This method could enable real-time brain analysis on mobile devices, revolutionizing clinical decision-making and telemedicine.}




\maketitle
\section{Introduction}\label{sec1}
Magnetic resonance imaging (MRI) technology is enhancing health care system by improving the diagnosis, prognosis, and treatment of diseases. This progress generates large amounts of data that clinicians and computational tools need to process. Over the past decade, Artificial Intelligence (AI) has been used with various objectives to assist physicians in leveraging all of this data \cite{AI_med}\cite{DL_med}.

Among the various applications of AI for MRI, one notable use is in the reconstruction of quantitative maps when no analytical model exists to describe the dependencies between the MR signal and the quantitative parameters. A specific example of this is magnetic resonance fingerprinting (MRF), a fast quantitative MRI technique that can produce multi-parametric maps in a single scan \cite{MRF_nature}. Initially, the obtained MR signal had to be compared against a precomputed or simulated signal dictionary to reconstruct the quantitative maps. This approach of using a dictionary faced significant challenges due to the 'curse of dimensionality,' which refers to the exponential growth in computational and memory requirements as the dimensionality, i.e. number of parameters, increases \cite{course_of_dim1} \cite{course_of_dim2}.\\

Recently, to overcome this issue, Barbieri et al. proposed the use of a fully connected Neural Network (NN) \cite{marco1} \cite{marco2}. The results of these studies demonstrated that the efficiency of NN was greater than or matched that of the original dictionary-based method in parameter map reconstruction. Additionally, NN exhibited better memory efficiency and reduced computational burden, indicating its substantial potential to address large-scale MRF challenges \cite{marco2}.

Although the NN offers rapid parameter estimation once trained, its training demands substantial computational resources and time, specifically due to the large number of images required.
In recent years, field-programmable gate array (FPGA) acceleration for NN algorithms has become a new direction \cite{FPGA_NN}, as it exhibits high throughput and low latency. After a traditional software validation, the NN algorithm can be transformed into a hardware version compatible with modern FPGAs, aimed at accelerating processing times by a factor of a few times to hundreds.\\

A recent work by Xiong et al. \cite{FPGA_brain_tumore_seg} developed an FPGA-accelerated NN for brain tumour segmentation in structural MRI images, with the aim of reducing the high computational time required by existing computer-aided detection systems. In contrast to conventional computing platforms, the FPGA accelerator presented by the authors exhibited significant improvements in speed and energy efficiency. 

Analysing patient data in real time during medical procedures can offer crucial diagnostic insights that greatly enhance the likelihood of success. According to Sanaullah et al. \cite{real_time_FPGA}, a low-latency Multi-Layer Perceptron (MLP) processor was developed using FPGA.  Their FPGA design surpasses both CPU and GPU implementations, achieving an average speedup of 144x and 21x, respectively.\\

These studies demonstrate the significant potential  but the NN is initially trained using software and subsequently implemented on the FPGA for inference purposes only. In the literature, there is an absence of training algorithms for NNs on FPGA being implemented, as this is an iterative process that is done once and offline \cite{FPGA_brain_tumore_seg}. 

However, MRF has a significant disadvantage: it is a non-standardised MR technique. Various factors, such as the scanner manufacturer, the magnetic field strength, and the number of parameters to be retrieved, can affect how the data are acquired and, consequently, how the quantitative maps are reconstructed. Therefore, whenever any single parameter is changed, the NN must be retrained to adapt to the new case. The retraining procedure is both computationally demanding and time-consuming, creating a substantial obstacle to the effective application and standardization of MRF techniques across various platforms and clinical environments.\\

The purpose of this work involves hardware programming for an FPGA-accelerated NN training algorithm for the reconstruction of MR parameters ($\text{T}_1$ and $\text{T}_2$) using clinical MRF.
To test the ability to accelerate the training process on FPGA, the original NN  must first be redesigned, i.e., simplified and quantized, to meet the available resources of the hardware accelerator. This would result in an important reduction in training time and power consumption.



\section{Materials and Methods}\label{mat_and_met}

\subsection{Deep Neural Network}

\begin{figure}[h!]
\centering
\includegraphics[scale=0.6]{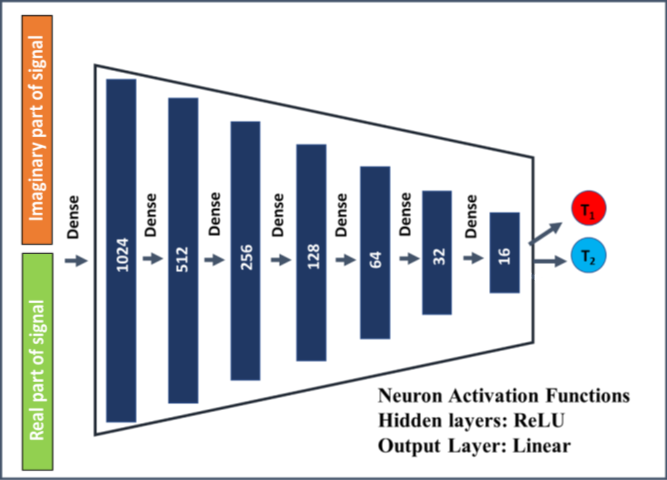}
\caption{Original NN proposed by Barbieri et al. \cite{marco1}\cite{marco2}. The NN is a fully connected with the Rectified Linear Unit (ReLU) used as a neuron activation function for hidden layers, while a linear activation function has been used for the output layer. Blue boxes represent layers and the numbers inside indicate the number of neurons. Image adapted from \cite{marco2}.}
\label{fig: original network}
\end{figure}

The N model introduced by Barbieri et al. \cite{marco1}\cite{marco2} is a feedforward network comprising nine fully connected layers. The Rectified Linear Unit (ReLU) serves as the activation function for the neurons in the initial 8 layers, whereas a linear activation function is employed for the output layer. The NN processes the real and imaginary components of the complex signal acquired with the MRI scanner to generate the $\text{T}_1$ and $\text{T}_2$ quantitative maps. The architecture of the NN is presented in Fig. \ref{fig: original network}.
The NN training procedure was supervised, with the Mean Squared Error (MSE) as the loss function, and was executed over 500 epochs, each consisting of 1000 gradient steps, and the learning rate was set to $10^{-4}$. To optimise the model, Adam Optimiser \cite{adam_optimizer} was used. The NN application was developed using the Python package Keras with TensorFlow \cite{tensorflow} backend and was run on an AMD Ryzen 9 3900 CPU with a total training time of about $16$ hours.
\\

\begin{figure}[h!]
\centering
\includegraphics[scale=0.5]{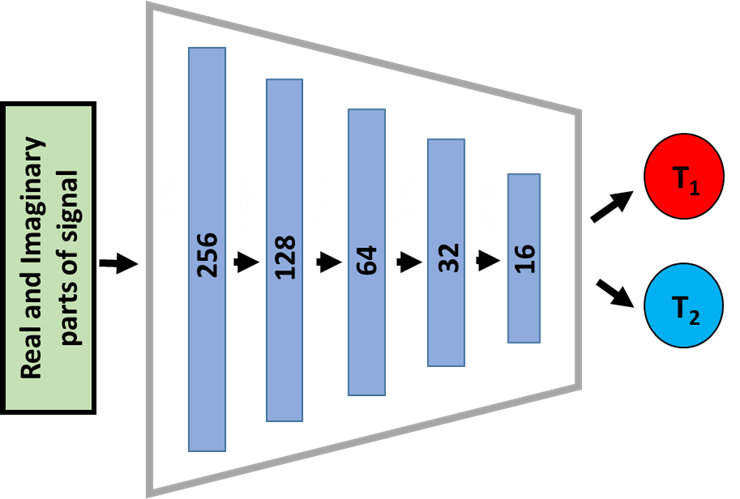}
\caption{Adapted NN obtained starting from the original NN. The first two layers were removed to reduce the number of parameters to be retrieved and the network is quantized using QAT to finally obtain a full-integer NN. Blue boxes represent layers and the numbers inside indicate the number of neurons.}
\label{fig: adapted network}
\end{figure}

To meet the available resources of the FPGA, the NN had to be adapted: the first two layers of the original NN were removed, reducing the number of parameters to be estimated. The network was then trained using the same dataset as the original network, consisting of 250M MRF simulated signals with different signal-to-noise ratio (SNR) and phase. To evaluate the performance, the trained NN was tested with 5000 never-before-seen synthetic signals. Finally, the NN was quantized using Quantization Aware Training (QAT) \cite{qat} so that the quantized model uses lower precision - i.e. all integer parameters - without affecting the NN performance. The adapted NN is shown in Fig. \ref{fig: adapted network}.

\subsection{FPGA Porting} 
The current status of firmware development for FPGA allows several approaches for the NN acceleration through hardware. A low-level HDL design approach, in which every firmware component is written in VHDL, without any high-level synthesis support, has been selected. This choice enables concurrent full structure control and data protection by exploiting an on-FPGA firewall security algorithm \cite{Grossi_2023}. 
The chosen FPGA board is the ALVEO U250 which disposes of 1.7M of Look-Up Tables (LUT), 3.4M of Flip-Flops (FF), 12k Digital Signal Processors (DSP) and 2.6k BRAMs.\\

To begin with, the function that represents the behavior of the single node, Eq. \eqref{eq: behaviour_node},  was implemented in VHDL, where $\sigma$ is the activation function, $x_i$ is the input to the node, $w_i$ is the weight associated to the link with the input $x_i$ and $b$ is the bias.
\begin{equation}
y = \sigma \left( \sum_{i=1}^{N_{inputs}} x_i \cdot w_i + b \right)
\label{eq: behaviour_node}
\end{equation}

Since Eq. \eqref{eq: behaviour_node} represents the behavior of all the nodes, it was generically implemented once and then used all the necessary times to cover all the node operations present in the NN.
To verify the proper functioning of the node, 16 nodes were completely deployed on the FPGA board. Identical inputs, weights, and biases were provided for both hardware and software implementations, and the results from Python and FPGA were subsequently compared.\\

Secondly, the backpropagation algorithm was implemented. As a starting point, the simple stochastic gradient descent was chosen, which describes how the parameters of the NN, i.e. weights and biases, are updated at each iteration during training. The implemented functions are shown in Eq. \eqref{eq: backprop}
\begin{equation}
\begin{aligned} 
  \delta^l &= \left( \left( \mathbf{w}^{l+1} \right)^T \delta^{l+1} \right) \circ \sigma'(z^l) \\ 
  \dfrac{\partial \mathcal{L}}{\partial \mathbf{w}^{l}} &= y^{l-1}\delta^l\, \, \, \, \, \, \, \text{and}\, \, \, \, \, \, \, \dfrac{\partial \mathcal{L}}{\partial b^l} = \delta^l
\end{aligned}
\label{eq: backprop}
\end{equation}
where $\mathcal{L}$ is the loss function, $l$ represents the layer number within the NN, $\mathbf{w}$ the matrix of weights between layer $l$ and $l+1$, $b^l$ the array of biases of the nodes of layer $l$, $\sigma '$ the derivative of the activation function and $z$ its argument, representing the weighted sum in Eq. \eqref{eq: behaviour_node}. \\

Finally, an assessment of the necessary resources was conducted taking into account the operation of the nodes, the backpropagation process, and the memory required to store weights and biases. This assessment comprises the number of LUTs, DSPs, and FFs needed for the NN to function.

\section{Results and Discussion}\label{result_disc}
Tab. \ref{tab: error_orig_quant} reports the error metrics both in the cases of the original and the quantized NN, showing that the quantization process did not affect the NN performance but simply reduced its precision, allowing it to work with full integer numbers.
\begin{table}[h]
  \centering
  \begin{tabular}{|c|c|c|c|c|}
    \hline
     & \multicolumn{2}{c|}{$T_1$} & \multicolumn{2}{c|}{$T_2$} \\ \hline
    & Original & Quantized & Original & Quantized \\ \hline
    MAPE (\%) & $2.15$ & $2.36$ & $8.89$ & $11.07$ \\ \hline
    MPE (\%) & $-0.66$ & $0.12$ & $0.02$ & $-3.12$ \\ \hline
    RMSE (ms) & $75$ & $78$ & $145$ & $148$ \\ \hline
  \end{tabular}
  \caption{Comparison between the error metrics in the case of the original and quantized NN. Results show good performance of the quantized NN.}
  \label{tab: error_orig_quant}
\end{table}

The synthesis tool provided by Vivado enables the transformation of high-level design descriptions into gate-level models, aiding in the development of efficient and optimised digital circuits. By analysing synthesis results, one can estimate the maximum achievable clock frequency for the implemented design. In our scenario, based on the synthesis outcomes of a single node and the backpropagation process, a clock frequency of $200$ MHz is totally feasible, with the possibility of increasing it to $250$ MHz.
Based on the estimation of the required resources for executing all the necessary operations on the FPGA, the whole network and backpropagation algorithm cannot be implemented on the FPGA but, it is feasible to implement 16 nodes of the second layer on the FPGA, as well as the backpropagation between the layers containing 16 and 32 nodes. Thus, by iterating these two blocks multiple times in a semiparallelised way, all the operational requirements of the network can be covered.\\

To verify the correct implementation of the single node function in VHDL, the outputs produced on both the FPGA and in software were compared after providing identical inputs, weights, and biases. The comparison produced promising results, as there was no difference between the Python outputs and those of the FPGA, indicating that the mathematical operations were correctly translated into VHDL.\\

The estimate of the necessary FPGA resources was $145$k LUTs, $5$k DSPs, and $146$k FFs. This implies that the entire NN and backpropagation use $8\%$ of the available LUTs and $40\%$ of the available DSPs, demonstrating that the algorithm's implementation is entirely viable from the resource point of view. PCI Express technology was chosen to communicate from the PC's CPU to the FPGA and back resulting in additional resources of $83$k LUTs, $148$kn FFs and $150$ BRAMs, the internal RAM memories of the FPGA. \\

Finally, a fairly accurate estimate of the training time can be made. Each node needs 4 clock cycles to perform its operations. The 16 nodes implemented on the FPGA work in a semi-parallel way, resulting in 56 clock cycles required for all levels. Similarly, the single backpropagation module requires 3 clock cycles, iterating through the entire process for a total of 104 clock cycles.  With a clock frequency of $200$ MHz, the clock period is $5$ ns, considering 250M training data the total training time results in: 
\begin{equation}
    \left(5\cdot\left(250'000'000\cdot\left(56+104\right)\right)\right)=200\, \text{s}
\end{equation}

This result highlights how the NN can be trained on FPGA in less than $5$ minutes, which is 200 times faster than the corresponding training on CPU. The proposed method poses a big step in the direction of real-time and personalized healthcare, opening the possibility of having an integrated NN hardware accelerator for map reconstruction inside the MRI scanner.

\section{Conclusion}\label{conclusions}
MRF is an advanced method for generating quantitative brain maps from a single scan, enhancing personalized healthcare. The reconstruction of these maps relies on artificial intelligence, which offers fast and accurate results once the NN is trained, although the training process itself is quite demanding. This study aims to develop a NN hardware accelerator, implementing both the inference and training stages on FPGA. Preliminary calculations indicate that with a clock frequency of $200$ MHz, the network can be trained in less than $5$ minutes. Future goals include the implementation of optimizing algorithms for addition and multiplication pipelines, clock domain management, and backpropagation efficiency improvement by developing an optimiser.
To make placement and parallelisation easier, the use of resources between LUTs and DSPs should be balanced at 24\% (i.e., add 274k LUTs to remove approximately 2k DSPs). This approach would be ideal for implementing the NN twice on an FPGA to achieve parallel processing.
By incorporating hardware NN accelerators into MRI scanners, this novel method has the potential to enable real-time brain map reconstruction, thus furthering the progress of precision medicine.



\end{document}